\documentstyle[NATO,numreferences,epsfig]{CRCKAPB}

\begin{opening}
\title{IDENTIFICATION OF COOL WHITE DWARFS IN
THE NOAO DEEP WIDE--FIELD SURVEY}
\author{M. Kilic}
\author{D.E. Winget}
\author{T. von Hippel}
\institute{The University of Texas at Austin,
Astronomy Department,
1 University Station, C1400,
Austin, TX 78712, USA}
\author{C.F. Claver}
\institute{Kitt Peak National Observatory, NOAO, Tucson, AZ 85726, USA}
\end{opening}
\begin{document}
\section{Introduction}
The chronology of star formation is recorded in the white dwarf luminosity function (WDLF). White dwarf (WD) structure implies a relatively simple connection between WD luminosity and age. First attempts to exploit WDs as chronometers [3,4,6] showed that the WDLF was a map of the history of star formation in the disk, and a significant shortfall of low-luminosity degenerates--the inevitable consequence of the finite age of the disk. The shortfall near log$(L/L_{\odot})$$\approx$$-$4.5 implies a disk age of 6.5--9.5 Gyr [2]. The WDLF from wide common proper motion binaries [5] does {\it not} show the shortfall seen by Liebert {\it et al.} (1988). This suggests that the disk is at least
$\sim$10.5 Gyr old. Both Liebert {\it et al.} (1988) and Oswalt {\it et al.} (1996) WDLFs were derived from proper motion catalogs, hence may be affected by significant kinematical bias as well as incompleteness. The simple fact is, the faintest, age-dependent end of the WDLF is not yet reliably determined.

\section{Identification of White Dwarfs}
The broadband photometric system has a limited
capacity to distinguish metal poor subdwarfs from
cool WDs. In the absence of
significant line blanketing, both the subdwarfs and WDs have broadband
colors which closely approximate those of a blackbody. However, by
comparing the flux through a Mg absorption line--centered filter, e.g., DDO51, Claver (1995) has shown
that cool WDs separate from other field stars of similar
$T_{eff}$ (Fig 1a). This is because the majority of cool WDs have
essentially featureless spectra around $5150\AA$, where even subdwarfs
show significant absorption from the Mgb triplet and/or MgH.
The uncertainties in the WDLF can be reduced
by the extension of the sample of cool WDs
through
supplementing the existing depth-oriented surveys i.e. the NOAO Deep Wide-Field Survey.
This survey would be free from the kinematic biases mentioned above.

\vspace{0.1in}
\begin{minipage}{2.9in}
\hspace*{-0.4in}
\epsfig{figure=kilic2_f1.ps,height=2.9truein,angle=270}
\end{minipage}
\hspace*{-0.25in}
\begin{minipage}{2.1in}
\vspace{-0.1in}
\epsfig{figure=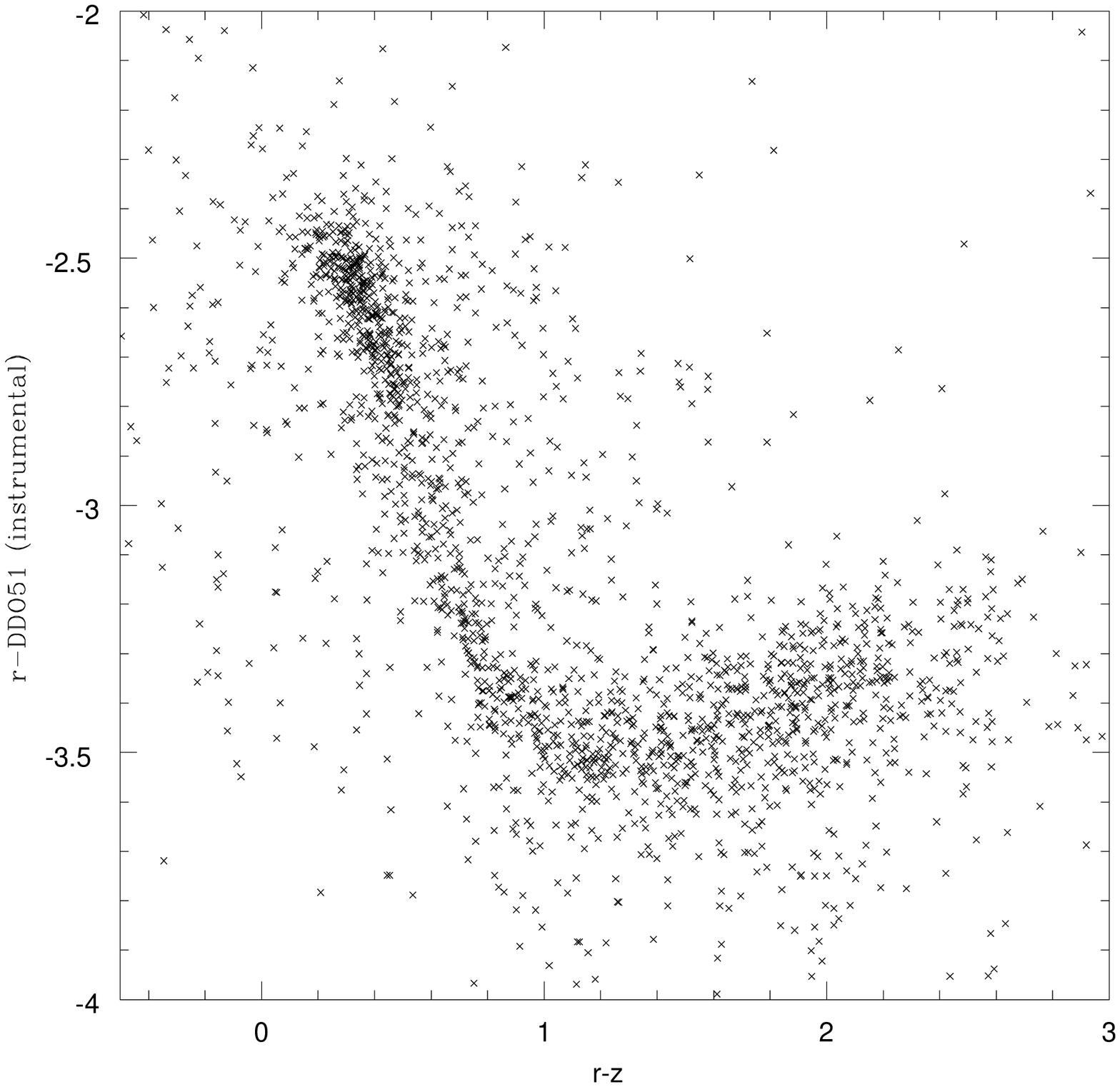,height=2.1truein,angle=0}
\end{minipage}

{\scriptsize {\it Figure1.} {\bf a)}Isolation of cool WDs in synthetic color-color diagram. {\bf b)} Observed diagram.} 
\vspace{0.15in}
\section{Observations and Results}
The NOAO Deep Wide-Field Survey is a deep optical and IR imaging survey of two 9 square degree fields. We have obtained DDO51 filter photometry of the northern survey field at Kitt Peak National Observatory in April 2002 using the 4m-Mayall Telescope and the MOSAIC imager. Figure 1b presents the observed color-color diagram for one of the fields.

A comparison of figures 1a and 1b shows that there are cool WD candidates in this field. These data, along with the follow-up spectroscopy with 9.2m Hobby-Eberly Telescope (HET), will ultimately resolve the controversies of the shape of the WDLF. Low resolution spectra for four WD candidates were obtained at HET using the LRS. We have discovered 2 cool WDs as companions to M dwarfs.

This work has been supported by NSF grant AST-9876730 and NASA grant NAG5-9321. M.K. acknowledges support from the AAS Travel Grant.

\end{document}